\documentclass[12pt, a4paper]{iopart}

\usepackage[a4paper]{geometry}
\usepackage[T1]{fontenc}\usepackage[latin1]{inputenc}
\usepackage{graphicx,color,booktabs,wrapfig,sidecap,floatflt,afterpage,cite,hvfloat}

\begin{document}

\pagestyle{plain}
	
\title{Adherence of the rotating vortex lattice in the noncentrosymmetric superconductor Ru$_{7}$B$_{3}$ to the London model}

\author{A. S.~Cameron$^{1}$, Y. V.~Tymoshenko$^{1,2}$, P.~Y.~Portnichenko$^{1}$, A. S. Sukhanov$^{1}$, M.~Ciomaga Hatnean$^{3,\ast}$, D.~McK. Paul$^{3}$, G.~Balakrishnan$^{3}$, R.~Cubitt$^{4}$, D.~S.~Inosov$^{1,5}$}
\address{$^{1}$ Institut f\"ur Festk\"orper- und Materialphysik, Technische Universit\"at Dresden, D-01069 Dresden, Germany}
\address{$^2$ J\"ulich Center for Neutron Science at MLZ, Forschungszentrum J\"ulich GmbH, Lichtenbergstra{\ss}e 1, 85748 Garching, Germany}
\address{$^{3}$ Department of Physics, University of Warwick, Coventry, CV4 7AL, United Kingdom}
\address{$^{4}$ Institut Laue-Langevin, 71 avenue des Martyrs, CS 20156, F-38042 Grenoble Cedex 9, France}
\address{$^5$ W\"urzburg-Dresden Cluster of Excellence on Complexity and Topology in Quantum Matter--ct.qmat, TU Dresden, 01069 Dresden, Germany}
\address{$^\ast$ Laboratory for Multiscale materials eXperiments, Research with Neutrons and Muons Division, Paul Scherrer Institute,  Villigen, 5232 Villigen PSI, Switzerland 
\\ 
Materials Discovery Laboratory, Department of Materials, Swiss Federal Institute of Technology Zurich, 8093 Zurich, Switzerland}

\begin{abstract}
The noncentrosymmetric superconductor Ru$_7$B$_3$ has in previous studies demonstrated remarkably unusual behaviour in its vortex lattice, where the nearest neighbour directions of the vortices dissociate from the crystal lattice and instead show a complex field-history dependence, and the vortex lattice rotates as the field is changed. In this study, we look at the vortex lattice form factor of Ru$_7$B$_3$ during this field-history dependence, to check for deviations from established models, such as the London model. We find that the data is well described by the anisotropic London model, which is in accordance with theoretical predictions that the alterations to the structure of the vortices due to broken inversion symmetry should be small. From this, we also extract values for the penetration depth and coherence length.
\end{abstract}

\date{\today}

\maketitle

\section*{Introduction}

Superconductivity in noncentrosymmetric (NCS) systems has attracted attention in recent years, following the prediction that the breaking of spatial inversion symmetry by the crystal structure leads to unconventional pairing symmetries with the mixing of singlet and triplet order parameters\cite{Sig91, Sig07, Gor01, Bau04, Fri04,  Bau12}. The first observed NCS superconductor was CePt$_3$Si~\cite{Bau04}, and since then their number has grown quite remarkably~\cite{Smi17}. The unconventional pairing state may lead to anisotropic gap structures and the emergence of accidental nodes~\cite{Hay06}. This, alongside the antisymmetric spin-orbit coupling that removes the spin degeneracy of the electronic bands~\cite{Rashba60}, results in the emergence of a plethora of unusual phenomena~\cite{Karki10, Chen11, Takimoto09, Samokhin04, Yua06}, including helical phases of the order parameter~\cite{Mineev94, Kau05} and spontaneous magnetisation at twin boundaries~\cite{Arahata13}. Many of these are of specific relevance to the vortex lattice (VL), such as the predicted stabilisation of the Fulde-Ferrel-Larkin-Ovchinnikov state~\cite{Tanaka07}, the emergence of the anomalous magnetoelectric effect, and a weakened paramagnetic limiting response~\cite{Edelstein95, Mineev05, Fujimoto05, Mineev11, Hiasa08}. The emergence of the magnetoelectric effect directly affects the VL, as there is a complex array of shielding currents associated with the flux lines. Several theoretical studies have been done in this regard, with predictions of an altered vortex structure~\cite{Yip05, Hay06} and the possibility of tangential components of the magnetic field~\cite{Mas06, Lu09, Kas13}. Nevertheless, these contributions are considered to be small and would be difficult to observe with conventional probes of condensed matter.

In our previous studies of the VL, we have observed not only the presence of singlet-triplet mixing~\cite{Cam22}, but also a highly unusual rotation of the VL with respect to the crystal lattice~\cite{Cam19}. Structural transitions of the VL are not simply common, having been observed in classical superconductors~\cite{Lav10}, cuprates~\cite{Gil02, Whi09, Whi14, Cam14}, pnictides~\cite{Furukawa11, Mor14}, and others~\cite{Esk97}, but should exist in all type-II superconductors \cite{Lav10}. Nevertheless, the change in the VL orientation we observed in Ru$_7$B$_3$ was not connected to a structural transition. Instead, when the field was applied along the \textbf{a} axis, the orientation of the VL at any field appeared not to be fixed but rather almost entirely dependent on the field history of the sample. Some examples can be seen in figure~\ref{Fig1}, which shows the VL at 0.2~T in panels (a) and (b), and at 0.4~T in panels (c) and (d), under different preparation conditions which will be discussed later. In our previous study, we proposed that the effects from both time-reversal symmetry breaking and the broken inversion symmetry may couple to the VL free energy and could drive a change in orientation. In this study, we investigate the VL form factor, to look for changes in vortex structure during the rotation. Both the unusual rotating behaviour of the VL, and the possibility of additional shielding currents mentioned in the previous paragraph, behoves investigation. However, we do not necessarily expect any deviations observable by neutron scattering, as the predicted effects should be small, on the order of Gauss. 

\section*{Experimental}

SANS measurements were performed on the D33 instrument at the Institut Laue Langevin in Grenoble, France~\cite{Dew08}. Incoming neutrons were velocity selected with a wavelength between 10 and 14~\AA, depending on the measurement, with a $\Delta \lambda / \lambda$ ratio of $\sim 10 \%$, and diffracted neutrons were detected using a position sensitive detector. The sample was mounted on a copper holder with the \textbf{a} and \textbf{c} directions in the horizontal plane, as detailed in our previous publication~\cite{Cam19}, and placed in a dilution refrigerator within a horizontal-field cryomagnet with the magnetic field applied along the neutron beam. The $T_{\rm c}$ of our sample was 2.4~K~\cite{Sin14, Cam22}, and since this was above the maximum stable temperature of the dilution refrigerator, the sample was cooled in zero applied field (zero field cooled, or ZFC) and the magnetic field was applied, and changed, while at base temperature. Measurements, such as those in figure~\ref{Fig1}, were taken by holding the applied field and temperature constant and rocking the sample throughout all the angles that fulfil the Bragg conditions for the first-order diffraction spots of the VL. Background measurements were taken in zero field and then subtracted from the in-field measurements to leave only the signal from the VL. Data reduction was performed with the GRASP software, and diffraction patterns were treated with a Bayesian method for handling small-angle diffraction data, detailed in Ref.~\cite{Hol14}.

The sample was an approximately cylindrical ingot of Ru$_{7}$B$_{3}$, with a length of 30~mm and a diameter of 5~mm. To bring the the \textbf{a} and \textbf{c} directions into the horizontal plane, the sample was mounted with the long axis of the cylinder near vertical, as illustrated in figure~1 of our previous publication~\cite{Cam19}. Therefore, it was not possible to illuminate all of the sample with neutrons, and so the neutron beam profile was selected to be a circle with a diameter approximately that of the sample. The illuminated sample volume could therefore be approximated by a Steinmetz solid.

\section*{Results}

\begin{figure} 
	\includegraphics[width=1\linewidth]{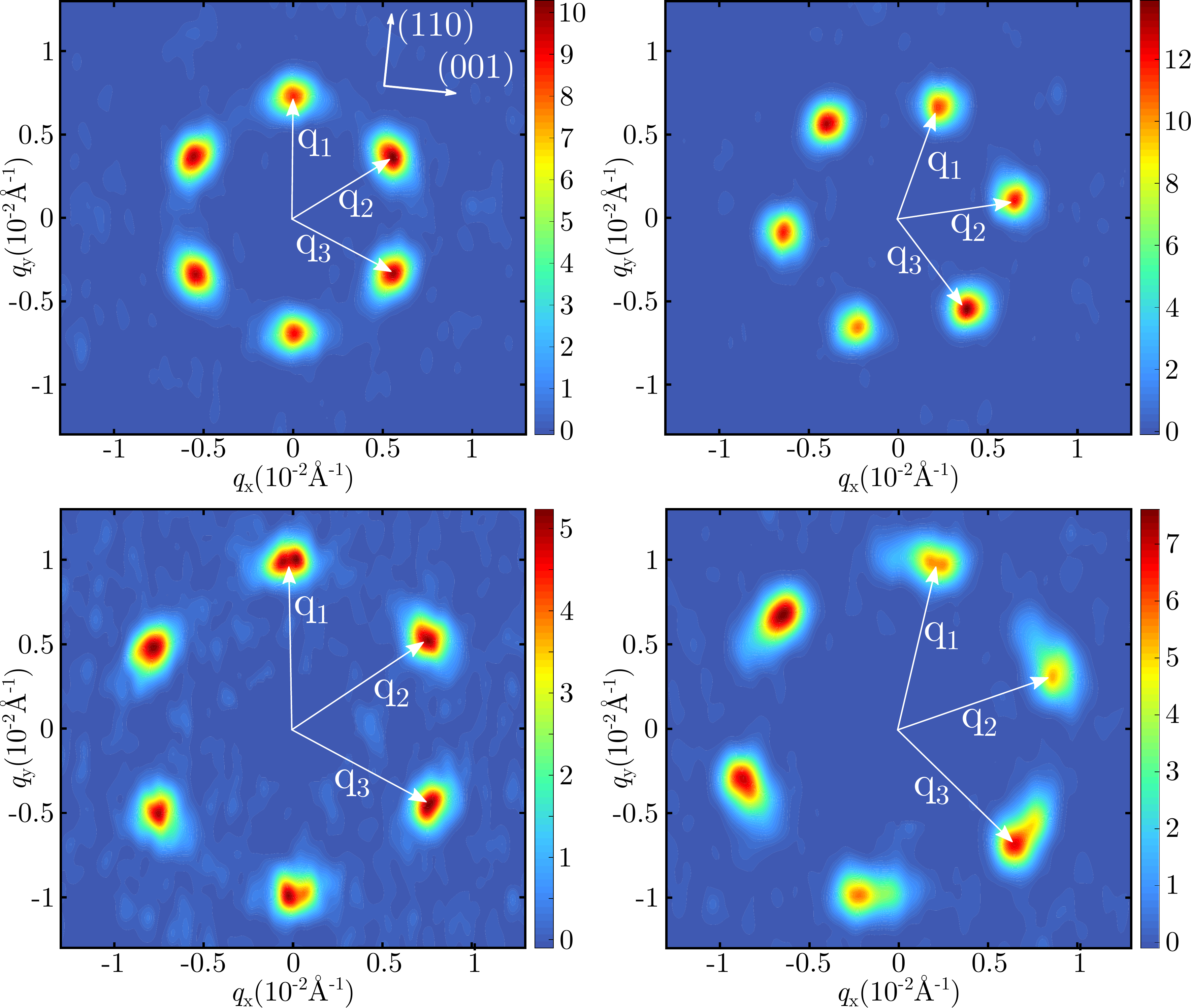}
	\caption{Diffraction patterns from the VL at several fields and different thermodynamic preparation paths. (a) VL at 0.2~T after ZFC. (b) VL at 0.2~T after decreasing the magnetic field from 1~T. (c) VL at 0.4~T after ZFC. (d) VL at 0.4~T after decreasing the field from 1~T.}
	\label{Fig1}
\end{figure}

Figure~\ref{Fig1} presents diffraction patterns from the VL for magnetic field applied parallel to the \textbf{a} axis, which are composed of the sum of all rocking angles which fulfilled the Bragg condition. It is in this orientation of the magnetic field that the field-history dependent rotation was observed~\cite{Cam19}. Panel (a) is a diffraction pattern at 0.2~T after the field was applied from zero at base temperature, and six first-order diffraction spots from a slightly distorted hexagonal lattice are observed. We can see that the orientation of the VL is slightly away from alignment with the crystal axes, which is a result of the rotation behaviour described previously. Panel (b) shows the VL after the field was increased to 1~T and then decreased back to 0.2~T, and we observe that the direction of the three reciprocal lattice vectors, $q_{1, 2, 3}$ have rotated away from their initial orientation. Panel (c) shows the VL at 0.4~T after applying the field from zero at base temperature, and panel (d) shows the same 0.4~T VL after the same field loop up to 1~T as was done for panel (b). 

The local field within the VL can be expressed as a sum over spatial Fourier components at the reciprocal lattice vector \textbf{q}. The magnitude of the Fourier component $F(\mathbf{q})$ is the form factor, and can be calculated from the integrated intensity, $I_{\mathbf{q}}$, of a VL Bragg reflection. This relation is given by~\cite{Chr77}:
\begin{equation}
I_{\mathbf{q}} = 2 \pi V \phi (\frac{\gamma}{4})^2 \frac{\lambda_{n}^{2}}{\Phi_{0}^{2} q} \vert F(\mathbf{q}) \vert^2,
\label{Eq1}
\end{equation}
where \textit{V} is the illuminated sample volume, $\lambda_n$ is the neutron wavelength, $\gamma$ is the magnetic moment of the neutron and $\Phi_0 = h / 2e$ is the flux quantum. The integrated intensity $I_{\mathbf{q}}$ was determined by fitting the rocking curves of the Bragg reflections to a Pseudo-Voigt line-shape~\cite{Ida00}, and this was corrected by the Lorentz factor, the cosine of the angle between the rocking axis and \textbf{q}~\cite{Squ96}.	

\begin{figure} 
\begin{center}
	\includegraphics[width=0.7\linewidth]{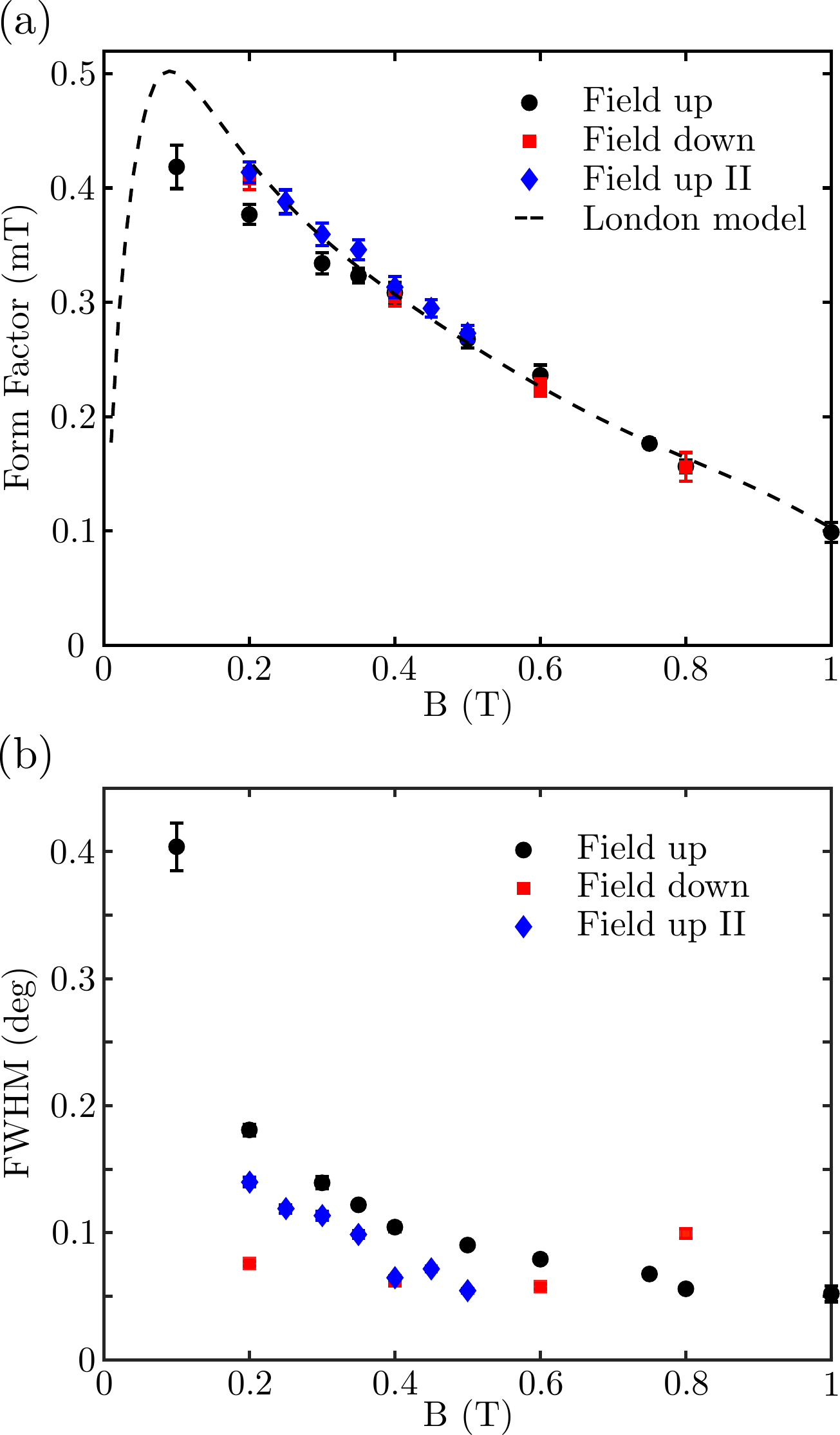}
	\caption{(a) Vortex lattice form factor as a function of applied magnetic field, for a series of scans described in the text, alongside a fit to the data from the anisotropic London model. (b) FWHM of the rocking curves from the same data as panel (a).}
		\label{Fig2}
\end{center}
\end{figure}

Figure~\ref{Fig2}(a) shows the average VL form factor for the first order diffraction spots illustrated in figure~\ref{Fig1}, as well as a fit to the anisotropic London model. There are three sets of data in this figure, each corresponding to a single scan up or down in magnetic field. The black circles correspond to an increasing field scan after the sample was cooled in zero field, going from 0.1~T to 1~T. The red squares are a decreasing field scan taken after the initial scan, starting from 0.8~T and descending to 0.2~T. The final scan, represented by blue diamonds, is an increasing field scan taken after the same field history that formed the decreasing field scan, going up to 0.5~T. These are the same data which formed the rotating vortex lattice scans of figure~2 in our previous investigation~\cite{Cam19}. 

Figure 2(b) shows the full-width at half maximum (FWHM) of the lineshape used to fit the rocking curves, from which the integrated intensity in equation~\ref{Eq1} was calculated. It contains the same three sets of data as in panel (a), using the same legend to identify them.

The vortex lattice is slightly distorted, indicating that the penetration depth and coherence length are not the same along the different crystallographic directions, meaning that it is appropriate to use the anisotropic London model, given in equation~\ref{LondonFF}. Since, however, the lattice is rotating, the contribution of each component of the penetration depth and coherence length along $q_{\rm x}$ and $q_{\rm y}$ will change as a function of field. We note, however, that the form factor for each species of Bragg spot characterised by the scattering vectors $q_{1}$, $q_{2}$, and $q_{3}$, are the same within error. We will therefore seek a single set of parameters for the anisotropic London model which reproduces the data for all three species of Bragg reflection simultaneously. 

There is one further consideration before we fit the data, which is the difference between the low field form factor of the initial field-up scan, and the same data of the other two scans. We see that the form factor of the initial scan is lower at low field than the following two scans, but that these scans all converge on the same values from around 0.4~T and above. Potentially, this could indicate a legitimate change in VL form factor, however, we note that the FWHM of the low field data in this scan, especially that at 0.1~T, is much larger than for the other data. Therefore, VL disorder which emerged from applying the field at low temperature may be affecting this data through the static Debye-Waller effect, which has been seen in VL studies before~\cite{Cam14}. Particularly, in the case of YBCO, it was observed that the form factor increased above a certain temperature, which was attributed to thermal energy forcing vortices out of pinning sites and thus reducing the static Debye-Waller effect. It is normally the case that pinning has the strongest effect on the vortex lattice at low field, and we suspect that increasing the field from zero while the sample was cold forced the vortex lattice through a more disordered region, and this disorder was only ``ironed out'' at higher fields where the inter-vortex interaction was much stronger. We therefore conclude that the static Debye-Waller effect is affecting these data, and do not include the low-field data up to 0.4~T from the ``field-up'' scan in our numerical analysis.

The anisotropic London model, which we use to fit the data, is given by the equation
\begin{equation}
F(\textbf{q}) = \frac{\langle B\rangle \exp(-c(q_{x}^{2}\xi_{b}^{2} + q_{y}^{2}\xi_{a}^{2}))}{q_{x}^{2}\lambda_{a}^{2} + q_{y}^{2}\lambda_{b}^{2}},
\label{LondonFF}
\end{equation}
where $\langle B \rangle$ is the average internal induction, $\xi_{i}$ is the coherence length along axis $i$, $\lambda_{i}$ is the penetration depth arising from supercurrents flowing in direction $i$, and $q_{x}$, $q_{y}$ are the in-plane Cartesian components of the scattering vector. The parameter \textit{c} accounts for the finite size of the vortex cores, and a suitable value for $c$ in our field and temperature range is 0.44 \cite{Cam22}. To fit this data, we will separate the data into the three different species of Bragg spot, denoted by the scattering vectors $q_{1,2,3}$ in figure~\ref{Fig1}, where each Bragg spot shares the components of the scattering vector, $q_{\rm x}$ and $q_{\rm y}$, with the corresponding reflection on the other side of the diffraction pattern. Once fitted, these will be averaged to return a single value of $F(\textbf{q})$ which can be compared to the data. It is important to note that during this procedure we checked to make sure that the fitted values of the form factor for each species of Bragg spot were equivalent to the other two within error, and this was the case.

We can see from figure~\ref{Fig2} that the model is a good fit for the data. It returned values of the penetration depth as $\lambda_{110} = 253 \pm  7.57$~nm, $\lambda_{001} = 244 \pm 8.49$~nm and the coherence length as $\xi_{110} = 13.7 \pm 0.65$~nm, $\xi_{001} = 9.8 \pm 0.66$~nm. These are mostly within the range of previously reported values of $\lambda = 214$~nm and $\xi=17.3$~nm~\cite{Fan09}, and $\lambda \parallel [100] = 311$~nm, $\lambda \parallel [001] = 352$~nm, $\xi \parallel [100] = 14.4$~nm and $\xi \parallel [001] = 13.8$~nm~\cite{Kas09}, although our values for the coherence length are a little shorter. In order to reproduce the fitting line in figure~\ref{Fig2}, the $q_{\rm x}$ and $q_{\rm y}$ values of each species of Bragg spot for each scan were fitted to a polynomial, which is a normal procedure when displaying the fit in these figures, and then averaged.

\section*{Discussion}


The data in figure~\ref{Fig2} is well represented by the anisotropic London model of equation~\ref{LondonFF}, for scans in both increasing and decreasing field, outside of the region where we believe the static Debye-Waller effect may be suppressing the measured intensity from neutron scattering. This indicates that although the vortex lattice is engaging in the unusual behaviour of rotating with respect to the crystal axes, in a manner not predicted by any prevailing theory of the vortex lattice, its gross structure remains the same as in other superconductors. In our previous work on Ru$_7$B$_3$, we proposed a mechanism whereby the effects of the broken inversion symmetry and time-reversal symmetry breaking could couple to the orientation of the vortex lattice, perhaps allowing for the field-history dependent orientation we observed. Our observation here, while not an indication of this mechanism, is in general agreement with this theory, as the predicted effects would be small.

The observation of London-type vortex behaviour is also in agreement with theories of the vortex lattice in NCS superconductors, which predict additional components of magnetic field to arise within the VL due to the magnetoelectric effect. These field components are expected to be small, on the order of a Gauss~\cite{Lu09}, and should not therefore alter the form factor as measured by neutron scattering, which is concurrent with what we observe here.

\section*{Conclusions}

We have measured the vortex lattice form factor in Ru$_7$B$_3$ for fields applied parallel to the \textbf{a} axis, during the phenomenon where the nearest neighbour directions dissociate from the crystal axes and have an orientation dependent on the field-history of the sample. We find that the data is well described by the anisotropic London model, which returns realistic values for the penetration depth and coherence length. This indicates that the rotation is not being driven by a large reshaping of the vortices, and also that the alterations of the vortex structure due to the broken crystal inversion symmetry are small. These observations are in agreement with the theoretical predictions to date.

\section*{Acknowledgements} 

This project was funded by the German Research Foundation (DFG) through the research grants IN 209/3-2 and IN 209/6-1, the Research Training Group GRK 1621, and by the Federal Ministry of Education and Research (BMBF) through the projects 05K16OD2 and 05K19OD1. A.S.S. acknowledges support from the International Max Planck Research School for Chemistry and Physics of Quantum Materials (IMPRS-CPQM). The work at Warwick was supported by EPSRC, UK, through Grant EP/M028771/1.
\\

\bibliographystyle{iopart-num}
\bibliography{Ru7B3}

\end{document}